\documentstyle[aps,preprint]{revtex}
\begin{document}
\draft
\def\ao{\hat{a}}       
\def\coh{\vert \alpha \rangle}
\def\ncoh{\vert -\alpha \rangle}
\def\co{{\hat{a}^\dagger}}
\def\cot{\hat{a}^{\dagger 2}}
\def\aot{\ao^2}
\def\aok{\ao^k}
\def\cok{\hat{a}^{\dagger k}}
\def\opfn{f(\hat{n})}
\def\opffn{\uppercase{f}(\hat{n})}

\bibliographystyle{unsrt}

\title{Generation of even and odd nonlinear coherent states } 

\author{S. Sivakumar \\
      Laboratory and Measurements Section\\ 307, General Services Building\\
	  Indira Gandhi Centre for Atomic Research\\
	  Kalpakkam, INDIA- 603 102.\\
       }
\maketitle

\begin{abstract}

We show that a class of even and odd nonlinear coherent states, defined
as the eigenstates of product of a nonlinear function of the number operator
and the square of the boson annihilation operator, can be generated
in the center-of-mass motion of a trapped and bichromatically laser-driven 
ion.
The nonclasscial properties of the states are studied.
\\

\end{abstract}


        Coherent states are important in many fields of physics\cite{jrk,rjg}.
  Coherent states $\coh$, defined as the eigenstates of the harmonic
oscillator annihilation operator $\ao$, $\ao \coh = \alpha \coh$ \cite{gla},
 have statistical properties like the classical radiation field.  In a 
harmonic  oscillator potential the centre of the  coherent state wavepacket 
follows the classical trajectory. There exist states of the
electromagnetic field whose properties, like squeezing, higher order
squeezing, antibunching and sub-Poissonian statistics \cite{dfw,loud}, are
 strictly quantum mechanical in nature. These states are called  nonclassical
 states. The coherent states define the limit between the classical and
 nonclassical behaviour of the radiation field as far as the nonclassical
 effects are considered.

	A generalization of the coherent states was done by
 q-deforming the basic commutation relation $[\ao,\co] =1$\cite{qcs,mfl}. 
 A further generalization is to define states that are eigenstates of the
 operator $\opfn \ao$,
\begin{equation}
\opfn\ao\vert f,\alpha\rangle=\alpha\vert f,\alpha\rangle,\nonumber
\end{equation}
 where $\opfn$ is a operator valued funtion of the number operator 
$\hat{n} = \co\ao$. These eigenstates are called nonlinear coherent states 
and are nonclassical.  In the linear limit, $\opfn =1$, the nonlinear coherent
 states become the usual coherent states $\coh$.  The nonlinear coherent 
states were introduced, as f-coherent states, in connection with the study of 
the oscillator whose frequency depends on its energy\cite{manko1,manko2}.  
A class of nonlinear coherent states can be realized physically as the 
stationary states of the center-of-mass motion of a trapped ion\cite{filho1}.  
These nonlinear coherent states exhibit nonclassical features like squeezing 
and  self-splitting.  

        Superposition of coherent states gives rise to states with 
nonclassical properties.  An important case is the superposition of 
the coherent states $\coh$ and $\ncoh$ and the resultant states are 
eigenstates of the operator $\aot$\cite{dodonov}.   The symmetric
combination is the even coherent state(ECS), $\vert\alpha,+\rangle$, 
 and its number state expansion is
\begin{equation}
\vert\alpha,+\rangle=[\cosh{\vert\alpha\vert}^2]^{-1/2}\sum_{n=0}^\infty
{\alpha^{2n}\over{\sqrt{(2n)!}}}\vert 2n\rangle.
\end{equation}
The antisymmetric combination is the odd coherent state (OCS), $\vert\alpha
,-\rangle$, given by
\begin{equation}
\vert\alpha,-\rangle=[\sinh{\vert\alpha\vert}^2]^{-1/2}\sum_{n=0}^\infty
{\alpha^{2n+1}\over{\sqrt{(2n+1)!}}}\vert 2n+1\rangle.
\end{equation}
The ECS has a squeezing but no antibunching effect.  The OCS has
an antibunching effect but no squeezing effect\cite{hillery,xia}.Both ECS
and OCS have oscillatory photon number distribution.
These states can be generated by various schemes: propagation in Kerr
medium\cite{yurke,mecozzi}, micromaser cavity experiments\cite{slosser}, 
quantum nondemolition experiments\cite{song}
and motion of a trapped ion\cite{filho2}.  The even and odd coherent 
states can be interpreted as Schr$\ddot{\hbox{o}}$dinger cat states for 
appropriately large values of $\alpha$\cite{yurke}.

 The notion of even and odd coherent states has been generalised to the
 case of nonlinear coherent states\cite{mancini,siva}. 
The even and odd nonlinear coherent states are defined as 
the eigenstates of the operator $\opffn\aot$, where $\opffn$ is a 
operator valued function of the number operator $\hat{n}$.  We denote the 
eigenstates as
$\vert \alpha,F\rangle$ and they satisfy
\begin{equation}\label{nlcsdef}
\opffn\aot \vert \alpha,F\rangle = \alpha\vert \alpha,F\rangle,
\end{equation}
where $\alpha$ is complex.  The above equation gives rise to the recurrence
relation
\begin{equation}\label{coeff}
\langle n+2\vert \alpha,F\rangle = \alpha {\langle n\vert \alpha,F\rangle
\over{F(n)\sqrt{(n+1)(n+2)}}},
\end{equation}
for $n=0,1,2,3...$
and the funtion $F(n)$ is obtained by replacing the number operator
 $\hat{n}$ in $\opffn$ by the integer $n$.  
The complex numbers $\langle n+2 \vert \alpha,F\rangle~~ (n=0,1,2,...)$ are 
the expansion coefficients of the state $\vert \alpha,F\rangle$ in the 
harmonic oscillator basis. The above recurrence relation between the expansion
 coefficients yields
\begin{eqnarray}
\langle 2n\vert \alpha,F\rangle&=&\alpha^n {\langle 0\vert \alpha,F\rangle\over
{F(2(n-1))!!\sqrt{(2n)!}}},\\
\langle 2n+1\vert \alpha,F\rangle&=&\alpha^n{\langle 1\vert \alpha,F\rangle
\over{F(2n-1)!!\sqrt{(2n+1)!}}},
\end{eqnarray}
where $F(2(n-1))!!=F(0)F(2)F(4)...F(2(n-1))$ and 
$F(2n-1)!!=F(1)F(3)F(5)...F(2n-1)$.  The function $F(k)!!$ is set 
equal to unity if the argument $k$ is less than or equal to zero.
The above relations yields all the 
coefficients, $n=1,2,3...$, in terms of $\langle 0\vert\alpha,F\rangle$ and
$\langle 1\vert\alpha,F\rangle$. The coefficients
$\langle 0\vert\alpha,F\rangle$ and $\langle 1\vert\alpha,F\rangle$ are fixed
by normalizing the state $\vert\alpha,F\rangle$. 
If we choose $\langle 1\vert\alpha,F\rangle=0$, the state $\vert\alpha,F\rangle$
 involves the superposition of even number(Fock) states and represents the
even nonlinear coherent state.  If $\langle 0\vert\alpha,F\rangle=0$,
the state $\vert\alpha,F\rangle$,  the superposition of odd number states, 
is the odd nonlinear coherent state.  We denote the even nonlinear coherent 
state as $\vert\alpha,F,+\rangle$ and the odd nonlinear coherent state as 
$\vert\alpha,F,-\rangle$. In the linear limit, $\opffn =1$, the even nonlinear
coherent state becomes the even coherent state and the odd nonlinear coherent
state becomes the odd coherent state.  Depending on the form of $\opffn$
the even and odd nonlinear coherent states may exhibit many of the 
nonclassical features.  It is interesting to note that the squeezed vacuum 
and the squeezed first excited state of the harmonic oscillator can be 
interpreted as even and odd nonlinear coherent states respectively.
The squeezed vacuum is the even nonlinear coherent state when 
$F(\hat{n})=1/(1+\co\ao)$ and the squeezed first excited
state is the odd nonlinear coherent state with $F(\hat{n})=1/(2+\co\ao)$.

        One of the interesting systems in quantum optics is the harmonically
trapped and laser-driven ion wherein the interaction between the ion and the
laser has nonlinear $\hat{n}$ dependence. This system has been studied 
in very many contexts:  nonlinear coherent states\cite{filho1},
vibronic Jaynes-Cummings interaction\cite{blockley},
nonlinear Jaynes-Cummings interaction\cite{vogel1}, generation of even and odd coherent
states\cite{filho2}, quantum signatures of chaos\cite{breslin}, quantum
nondemolition measurements\cite{filho3}, quantum logic operations\cite{monroe},
 engineering of Hamiltonian\cite{filho4}
and generation of ampltiude-squared squeezed states\cite{zeng}.
 In the present contribution we show that a class of even and odd
nonlinear coherent
states can be generated in the center-of-mass motion of a harmonically trapped
ion via bichromatic laser excitation.  We also study the  nonclassical 
properties of the states produced. 

	We consider a single ion, having an electronic transition frequency
$\omega$ and a lower(second) vibrational sideband with respect to that
frequency,  trapped in harmonic potential of frequency $\nu$.  
Two laser fields, tuned, repectively, to $\omega$ and the vibrational sideband 
transition frequency interact with the ion.  The Hamiltonian of this system 
in the optical rotating-wave approximation can be written as\cite{filho2}
\begin{equation}
\hat{H} =\hat{H}_0+\hat{H}_{int}(t),
\end{equation}
with
\begin{equation}
\hat{H}_0=\hbar\nu \co\ao + \hbar\omega\hat{\sigma}_{22}.
\end{equation}
The free Hamiltonian $\hat{H}_0$ describes the free motion of the internal 
and external degrees of freedom and the interaction Hamiltonian 
$\hat{H}_{int}$,
\begin{equation}
\hat{H}_{int}(t)=\lambda[E_0 e^{[-i(k_0\hat{x}-\omega t)]}+E_1
 e^{[-i(k_1\hat{x}-(\omega-2\nu) t)]}]\hat{\sigma}_{12} + H.c.,
\end{equation}
 describes the interaction of the ion with the two laser fields.
The operators $\hat{\sigma}_{ij}~~(i,j=1,2)$ are the electronic 
flip operators correspodning to the transition  $\vert j\rangle\rightarrow
\vert i\rangle$ and $\ao$ is the annihilation operator for the vibrational
motion of the ion in the harmonic potential.  The constant $\lambda$ is the 
electronic coupling matrix element and $k_0,k_1$ are the wave vectors of the
laser fields.  The operator of the center-of-mass position of the ion is
\begin{equation}
\hat{x} = {\eta\over{k_L}}(\ao+\co),
\end{equation}
where $\eta$ is the Lamb-Dicke parameter and $k_L\simeq k_0\simeq k_1$ is 
the wave vector of the driving laser field.  

	In the interaction picture, defined by the unitary transformation
$\exp(-{{i\hat{H}_0t}\over{\hbar}})$, the interaction Hamiltonian becomes
\begin{eqnarray}
\hat{H}'_{int}&=&\exp(-{{i\hat{H}_0t}\over{\hbar}})\hat{H}_{int}
\exp({{i\hat{H}_0t}\over{\hbar}})\\
&=&\hbar\Omega_1 \exp(-\eta^2/2)\hat{\sigma}_{12} \left[\sum
_{k,l=0}^\infty {{(i\eta)^{k+l}}\over{k!l!}}e^{i(k-l-2\nu)t}\cok
\ao^l 
+{\Omega_0\over{\Omega_1}}
\sum_{k,l=0}^\infty{{(i\eta)^{k+l}}\over{k!l!}}e^{i(k-l)t}\cok\ao^l\right] + 
H.c.
\end{eqnarray}
where $\Omega_{i}={\lambda E_{i}\over{\hbar}}~~(i=1,2)$ are the Rabi frequency
of the two laser fields tuned to the electronic transition and the second 
sideband respectively.  In the rotating wave approximation, neglecting terms
rotating with frequencies $\nu$ or more, the interaction picture Hamiltonian
becomes
\begin{equation}
\hat{H}'_{int}=\hbar\Omega_1 \exp(-\eta^2/2)\hat{\sigma}_{21}
\hat{F} + H.c.,
\end{equation}
with
\begin{equation}\label{defnf}
\hat{F}=\sum_{k=0}^\infty{{{(i\eta)^{2k+2}}\over{k!(k+2)!}}\cok\ao^{k+2}}
+{\Omega_0\over{\Omega_1}}
{\sum_{k=0}^\infty{{{(i\eta)^{2k}}\over{k!^2}}\cok\ao^k}}.
\end{equation}

        The time evolution of the system is governed by the master equation
for the vibronic density operator $\hat\rho$,
\begin{equation}\label{master}
{d\over{dt}}\hat{\rho} = -{i\over{\hbar}}[\hat{H}_{int}',\hat{\rho}]+
{\Gamma\over 2}(2\hat{\sigma}_{12}\hat{\rho'}\hat{\sigma}_{21}-
\hat{\sigma}_{22}\hat{\rho}-\hat{\rho}\hat{\sigma}_{22}),
\end{equation}
where the second term is to include the effect of spontaneous emission with 
energy
relaxation rate $\Gamma$ and 
\begin{equation}
\hat{\rho'} = {1\over 2}\int_{-1}^{1}dy W(y)e^{i\eta(\ao+\co)y}\hat{\rho}
e^{-i\eta(\ao+\co)y},
\end{equation}
accounts for changes in the vibrational energy due to spontaneous emission.
$W(y)$ gives the angular distribution of spontaneous emission.
  The steady state solution
$\hat{\rho}_s$ of Eq. (\ref{master}) is obtained by setting ${d\over{dt}}
\hat{\rho}=0$.  If we make the ansatz that $\hat{\rho}_s$ is given by
\begin{equation}
\hat\rho_s = \vert 1)\langle\zeta\vert\langle\zeta\vert ( 1\vert,
\end{equation}
where $\vert 1)$ is the electronic ground state and $\vert\zeta\rangle$
is the vibrational state of the ion, then the state $\vert\zeta\rangle$ obeys
\begin{equation}
\hat{F}\vert\zeta\rangle = 0.
\end{equation}
Using $\hat{F}$ given by Eq. (\ref{defnf}) we get
\begin{equation}
\langle n+2\vert\zeta\rangle = {\Omega_0\over{\Omega_1\eta^2}}{{(n+1)(n+2)L_{n}
^{0}(\eta^2)}
\over{\sqrt{(n+1)(n+2)}L_{n}^{2}(\eta^2)}}\langle n\vert\zeta\rangle
\end{equation}
where $L_n^m$ is an associated Laguerre polynomial defined by
\begin{equation}
L_n^m(x)=\sum_{l=0}^n{\left(\begin{array}{c}{n+m}\\n-l\end{array} \right)}
{{(-x)^l}\over{l!}}.
\end{equation}
The numbers $\langle n+2\vert\zeta\rangle$ are the expansion 
coeffieients for the state $\vert\zeta\rangle$ in the Fock states basis.
Comparing with Eq. (\ref{coeff}) indicates that the state 
$\vert\zeta\rangle$ is an even or odd nonlinear coherent state with
\begin{eqnarray}
\alpha&=&{\Omega_0\over{\Omega_1\eta^2}}\label{defnal}\\
\noalign{\hbox{and}}\nonumber\\
F(n)&=& L_n^2(\eta^2)[(n+1)(n+2)L_n^0(\eta^2)]^{-1}\label{defnfln}.
\end{eqnarray}
In the limit $\eta\rightarrow 0$ the function $F(n)$ becomes $1\over 2$ for
all $n$.  Hence in the small $\eta$ limit the even and odd nonlinear coherent
states become the ECS and OCS respectively.

        If the initial state of the ion is a combination of even(odd) number
 states then the state of the system at later times will involve a 
superposition of even(odd) number states only as the master equation 
Eq. (\ref{master}) contains only even powers of $\ao$ and $\co$.  If the 
initial state of the ion is the vacuum state then the stationary state of the
 system is given by
\begin{equation}\label{enlcs}
\vert\alpha,F,+\rangle=N\sum_{n=0}^\infty{\alpha^n\over{\sqrt{(2n)!}}
F(2n-2)!!}\vert2n\rangle;\\
N^{-1}=\sqrt{\sum_{n=0}^\infty{\vert\alpha\vert^{2n}\over{(2n)!
(F(2n-2)!!)^2}}}
\end{equation}
where $\alpha$  and $F(n)$ are defined by Eq.(\ref{defnal}) and
Eq.(\ref{defnfln}) respectively.  This state is the even nonlinear coherent
state for the vibrational motion of the center-of-mass of the ion in the
harmonic potential.  The behaviour of the expansion coefficients 
$\langle n\vert\alpha,F,+\rangle$  is highly oscillatory  becoming zero for
odd $n$.  This oscillatory behaviour is one of the nonclassical features.

     The ECS exhibits squeezing in the $p-$quadrature which is defined as 
$i(\co-\ao)/\sqrt{2}$. For even nonlinear coherent states the expectaction 
values of 
$\ao$ and $\co$ become zero and the uncertainty in $p$ is given by
\begin{eqnarray}
\langle(\triangle\hat{p})^2\rangle&=&\langle\hat{p}^2\rangle-\langle\hat{p}
\rangle^2,\\
&=&{1\over{2}}\left[1+2\langle\co\ao\rangle-2\langle\aot\rangle\right],
\end{eqnarray}
where the expansion coefficients of the even nonlinear coherent state in the
harmonic oscillator basis are taken to be real.  In Fig. (1) we have shown the 
uncertainty in $p$ as a function of $\eta$ for the states defined by 
Eq. (\ref{enlcs}).  From Fig. (1) it is clear that the uncertainty in $p$ is 
less than that of the vacuum state value of 0.5 indicating that 
 the states exhibit squeezing. As $\eta$ increases the 
uncertainty in $p$ approaches that of the vacuum state.  The reason for this 
behaviour is the following. As $\eta$ increases the occupation number 
distribution $p(n)=\vert\langle n\vert\alpha,F,+\rangle\vert^2$ starts peaking 
near $n=0$.  To make this explicit we have shown in Fig. (2) the occupation 
number distribution $p(n)$ as a function of $n$ for various values of $\eta$.

        If the initial state of the ion is the first excited state of the 
harmonic trap then the state of the system at later times will involve only 
odd number states. The resultant stationary state of the system is an odd 
nonlinear coherent state  given by
\begin{equation}\label{onlcs}
\vert\alpha,F,-\rangle=N\sum_{n=0}^\infty{\alpha^n\over{\sqrt{(2n+1)!}}
F(2n-1)!!}\vert2n+1\rangle;~
N^{-1}=\sqrt{\sum_{n=0}^\infty{\vert\alpha\vert^{2n}\over{(2n+1)!
(F(2n-1)!!)^2}}},
\end{equation}
where $F(n)$ and $\alpha$ are again defined by Eq.(\ref{defnfln}) and 
Eq.(\ref{defnal}) respectively. As in the case of even nonlinear 
coherent states the behaviour of occupation number distribution  of the odd 
nonlinear coherent state, Eq.(\ref{onlcs}), is oscillatory becoming zero for 
even $n$. The occupation number distribution $p(n)$ of the odd nonlinear
coherent states is sub-Poissonian.  A state is said to exhibit sub-Poissonian
statistics if the $q$ parameter\cite{mandl}, defined as
\begin{equation}
q={{\langle\hat{n}^2\rangle-\langle\hat{n}\rangle^2}\over{\langle\hat{n}
\rangle}
}-1,
\end{equation}
is negative. Negative $q$ indicates that the state is nonclassical.  
For the coherent states $q$ is zero.  Fig. (3) shows the $q$ parameter as a 
function of $\alpha$ for the odd nonlinear coherent states of Eq. (\ref{onlcs}).
  It is clear that the states are sub-Poissonian.  It is interesting to note 
that the $q$ value for large values $\eta$ approaches the value of that of 
the first excited of the harmonic oscillator.  The reason being that the 
occupation number distribution gets concentrated at $n=1$ as $\eta$ 
increases.

	In conclusion, we have shown that a class of even and odd nonlinear
coherent states can be generated in the center-of-mass motion of a trapped
and bichromatically laser driven ion.  These even and odd nonlinear coherent
states are nonclassical.  The even nonlinear coherent state exhibits squeezing
while the odd nonlinear coherent states exhibits sub-Poissonian statistics.
Both the even and odd nonlinear coherent states have oscillatory occupation
number distribution.  

	The author acknowledges Drs. V. Balakrishnan,  M. V. Satyanarayana
and D. Sahoo for useful discussions.

\begin{figure}
\caption{Uncertainty S, $\langle p^2\rangle -\langle p\rangle^2$,  in $p$ as a
 function of $r$ for ${\Omega_0\over\Omega_1}=.001$(solid) and 
${\Omega_0\over\Omega_1}=.0001$(dashed)
for the state $\vert\alpha,F,+\rangle$.  $r$ represents real $\eta$.} 
\label{fig1}
\end{figure}
\begin{figure}
\caption{Occupation number distribution $p(n)$ as a function of $n$ for the
state $\vert\alpha,F,+\rangle$ for various $\eta$ values and 
${\Omega_0\over\Omega_1}=.0001$. (a) $\eta=.008$,
(b) $\eta=.012$, (c) $\eta=.02$, and (d) $\eta=.1$.
}
\label{fig2}
\end{figure}

\begin{figure}
\caption{Mandel's $q$ parameter as a function of $r$ for 
${\Omega_0\over\Omega_1}=.001$ for the state  $\vert\alpha,F,-\rangle$. 
$r$ represents real $\eta$.}
\label{fig3}

\end{figure}


\begin{thebibliography}{40}
\bibitem{jrk}{J.R. Klauder and B.-S. Skagerstam, {\it{Coherent States-
Applications
in Physics and Mathematical Physics}}  (World Scientific, Singapore 1985).}
\bibitem{rjg}{W.-M. Zhang, D.H. Feng and R. Gilmore, Rev. Mod. Phys. {\bf 62},
 867 (1990).} 
\bibitem{gla}{R.J. Galuber, Phys. Rev. {\bf130}, 2529 (1963); {\bf131}, 2766
 (1963); Phys. Rev. Lett. {\bf 10}, 84 (1963).}
\bibitem{dfw}{D.F. Walls, Nature(London), {\bf 306}, 141 (1983).}
\bibitem{loud}{R. Loudon and P.L. Knight, J. Mod. Opt. {\bf 34}, 709 (1987).}
\bibitem{qcs}{L.C. Biedenharn J.Phys.A {\bf 22}, L873 (1989).}
\bibitem{mfl}{A.J. Macfarlane, J. Phys. A {\bf 22}, 4581 (1989).}
\bibitem{manko1}{V.I. Man'ko, G.Marmo, F. Zaccaria and E.C.G. Sudarshan in
 Proceedings of the IV Wigner Symposium, edited by N. Atakishiyev, 
 T.Seligman
and K.B.Wolf (World Scientific, Singapore, 1996), pp. 421-428.}
\bibitem{manko2}{V. I. Man'ko, G. Marmo, F. Zaccaria and E. C. G. Sudarshan,
 Physica Scripta {\bf 55}, 528 (1997).}
\bibitem{filho1}{R.L. de Matos Filho and W. Vogel, Phys. Rev.A {\bf54}, 4560 
(1996)}.
\bibitem{dodonov}{V. V. Dodonov, I. A. Malkin and V. I. Man'ko, Physica
{\bf 72}, 597 (1974); V. Buzek and P. L. Knight, {\it{Quantum Interference,
 Superposition of Light and Nonclassical Effects}}, in "Progress in Optics", 
edited by E. Wolf (North-Holland, Amsterdam, 1995), Vol. 34}
\bibitem{hillery}{M. Hillery, Phys. Rev. A {\bf 36}, 3796 (1987).}
\bibitem{xia}{Y. Xia and G. Guo, Phys. Lett. A {\bf136}, 281 (1989).}
\bibitem{yurke}{B. Yurke and D. Stolers, Phys. Rev. Lett. {\bf57}, 13 (1986).}
\bibitem{mecozzi}
{A. Mecozzi and P. Tombesi, {\it{ibid.}} {\bf58}, 1055(1987); P. Tombesi and
A. Mecozzi, J. Opt. Soc. Am. B {\bf4}, 1700 (1987); G. J. Milburn and 
C. A. Holmes, Phys. Rev. Lett. {\bf56}, 2237 (1986); M. Wolinsky and 
H. J. Carmichael, {\it{ibid.}} {\bf60}, 1836 (1988).}
\bibitem{slosser}{J. J. Slosser and P. Meystre, Phys. Rev. A {\bf41}, 3867 
(1990); M. Wilkens and P. Meystre, {\it{ibid.}} {\bf43}, 3832 (1991).}
\bibitem{song}{S. Song, C. M. Caves, and B. Yurke, Phys. Rev. A {\bf41},
5261 (1990); M. Brune {\it{et al.}}, {\bf45}, 5193 (1992).}
\bibitem{filho2}{R. L. de Matos Filho and W. Vogel, Phys. Rev. Lett. {\bf76},
 608 (1996).} 
\bibitem{mancini}{S. Mancini, Phys. Lett. A {\bf233}, 291 (1997).}
\bibitem{siva}{S. Sivakumar, Phys. Lett. A (To appear).}
\bibitem{blockley}{C. A. Blockley and D. F. Walls, Phys. Rev. A {\bf 47}, 
2115 (1993); J. I. Cirac, R. Blatt, A. S. Parkins and P. Zoller, {\it ibid.}
 {\bf49}, 1202 (1994)}.
\bibitem{vogel1}{W. Vogel and D.-G. Welsch, Phys. Rev. A {\bf40}, 7113 (1989);
 W. Vogel and R. L. de Matos Filho, {\it ibid.} {\bf52} 4214 (1995).}
\bibitem{breslin}{J. K. Breslin, C. A. Holmes and G. J. Milburn, Phys. Rev. A
{\bf56}, 3022 (1997).}
\bibitem{filho3}{R. L. de Matos Filho and W. Vogel, Phys. Rev. Lett. {\bf76}
4250 (1996); L. Davidovich, M. Orszag and N. Zagury, Phys. Rev. A {\bf54}, 5118 
(1996).}
\bibitem{monroe}{C. Monroe, {\it{et al.}}, Phys. Rev. A {\bf55}, R2489 (1997).}
\bibitem{filho4}{R. L. de Matos Filho and W. Vogel, Phys. Rev. A {\bf58},
R1661 (1998).}
\bibitem{zeng}{H. Zeng, Phys. Lett. A {\bf247}, 273 (1998).}
\bibitem{mandl} {L.Mandel, Opt. Lett. {\bf 4}, 205(1979).}
\end{thebibliography}
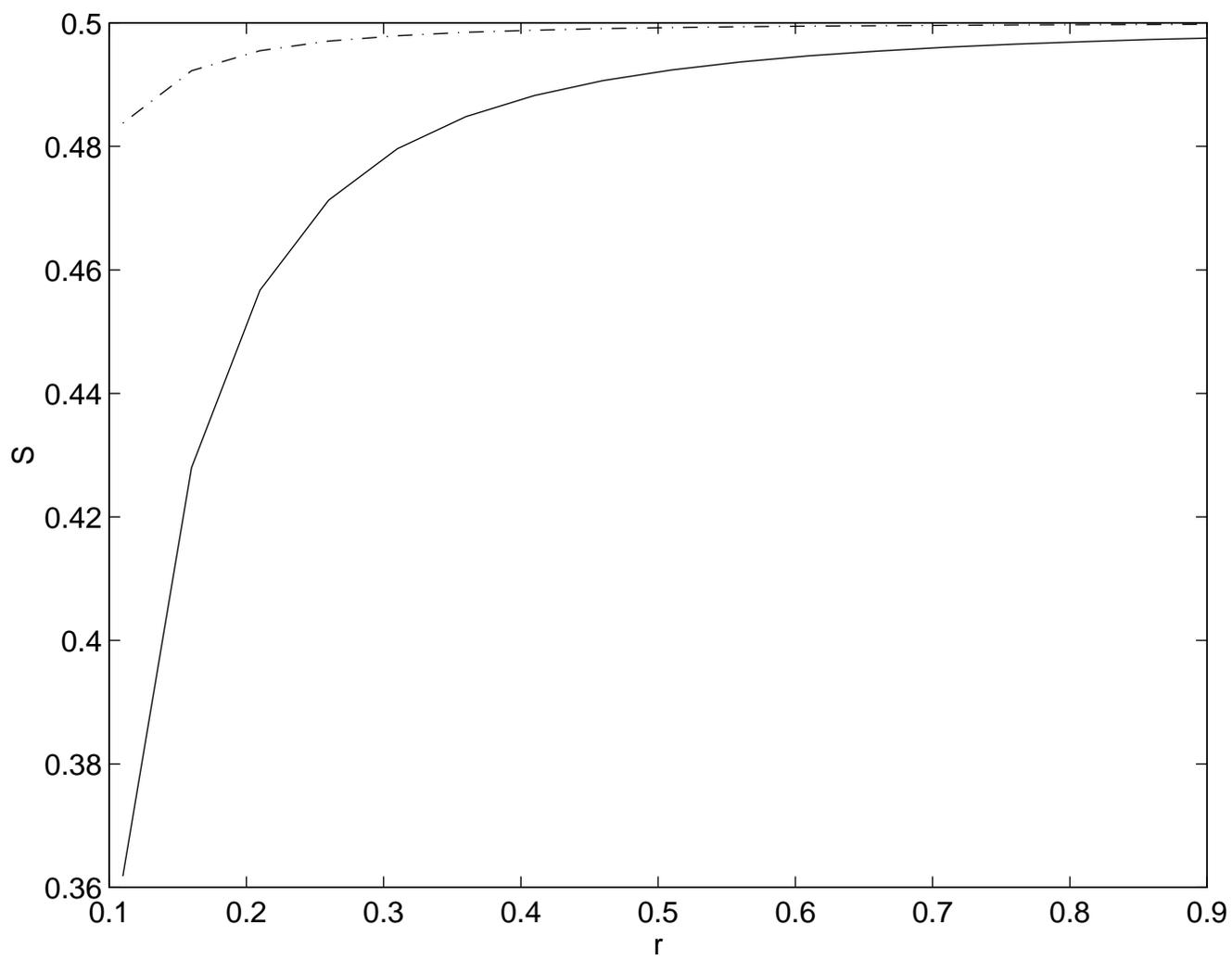
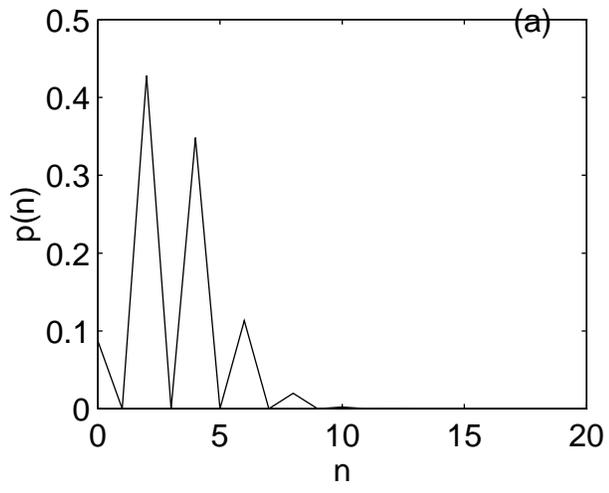
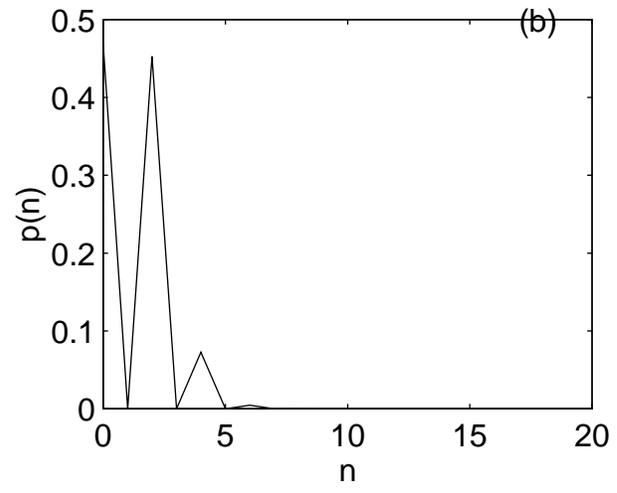
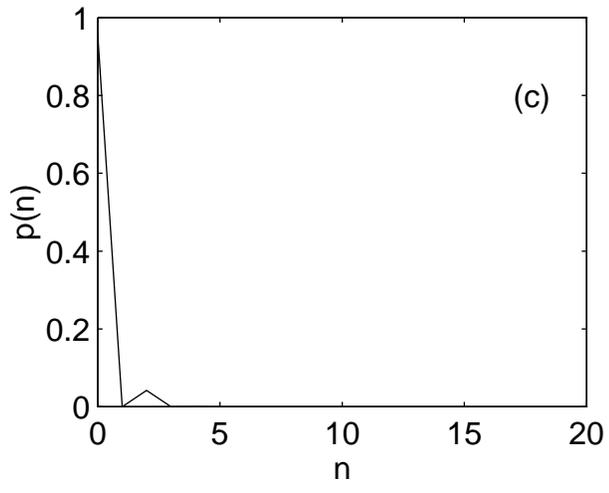
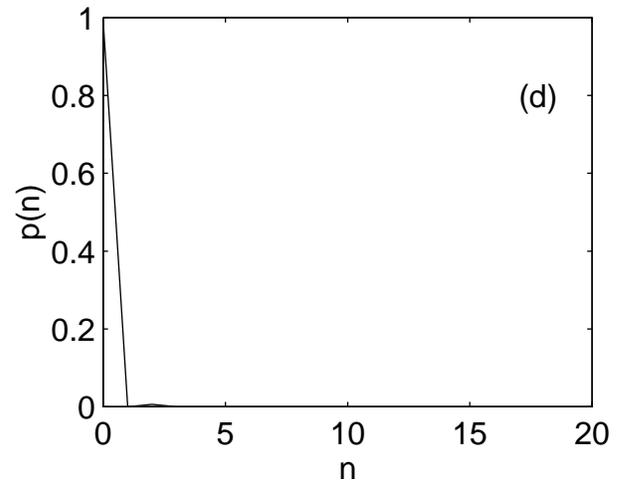
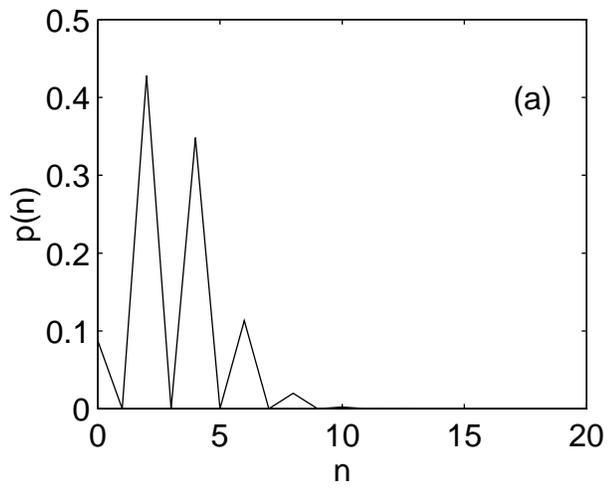
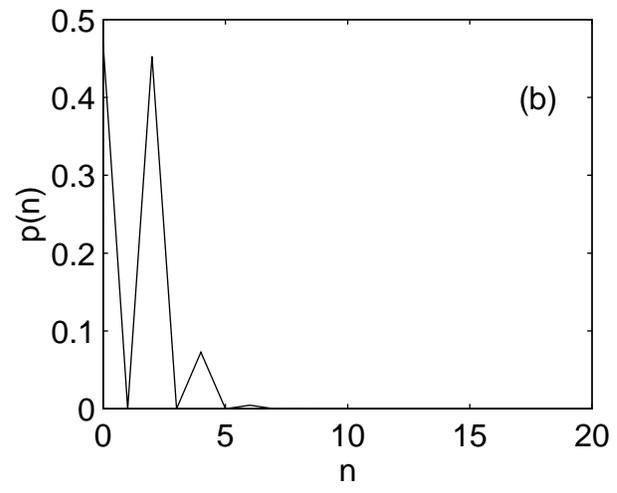
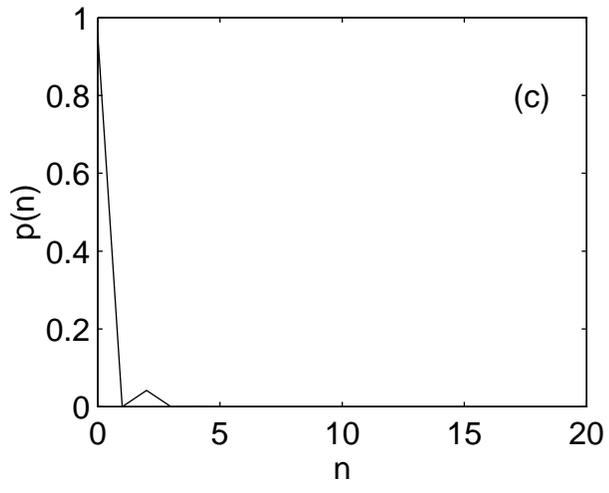
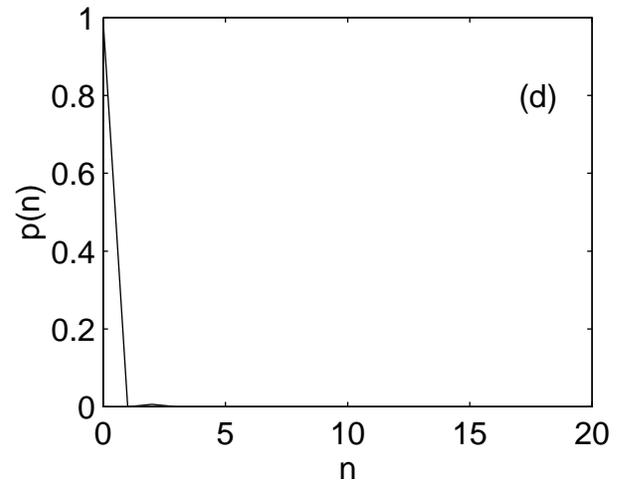
\end{document}